# A Blockchain-based Decentralized Data Sharing Infrastructure for Off-grid Networking

Harris Niavis  
*Yale University*  
New Haven, CT, USA  
charilaos.niavis@yale.edu

Nikolaos Papadis  
*Yale University*  
New Haven, CT, USA  
nikolaos.papadis@yale.edu

Leandros Tassiulas  
*Yale University*  
New Haven, CT, USA  
leandros.tassiulas@yale.edu

*Abstract*—Off-grid networks are recently emerging as a solution to connect the unconnected or provide alternative services to networks of possibly untrusted participants. The systems currently used, however, exhibit limitations due to their centralized nature and thus prove inadequate to secure trust. Blockchain technology can be the tool that will enable trust and transparency in such networks. In this paper, we introduce a platform for secure and privacy-respecting decentralized data sharing among untrusted participants in off-grid networks. The proposed architecture realizes this goal via the integration of existing blockchain frameworks (Hyperledger Fabric, Indy, Aries) with an off-grid network device and a distributed file system. We evaluate the proposed platform through experiments and show results for its throughput and latency, which indicate its adequate performance for supporting off-grid decentralized applications.

## I. Introduction

The recent emergence of decentralized technologies like distributed ledgers has gained significant attention and their potential accelerates the moving towards a decentralized ecosystem [1] that will shape the Web 3.0 and will enable the development of earlier unattainable peer-to-peer systems. As the current centralized Internet infrastructure is exposed daily by data breaches and identity thefts with huge financial, reputation and privacy costs [2], [3] for businesses and individuals, the need for the re-decentralization of the web - from infrastructure to protocols, applications and identities - is gradually realized.

Although this transformation has already started, we are still at an early stage inventing the building blocks that will allow us to build the future decentralized Internet on concrete foundations. The ongoing research is setting a scene analogous to the Internet Revolution in 1994 when all the network protocols were invented, which led to the later development of famous applications like Facebook and Netscape. Similarly today, experts are gathering across the world in workshops with the goal of designing and building the next generation of the Web [4].

Off-grid networking is part of this emerging ecosystem and started back in 80s and 90s, when communities created communication networks outside the main grid with the goal of connecting and socializing on their own terms and independently of service providers. These community networks evolved continuously until today, trying to keep up with the constant change of the technology, and got a lot of attention as a bottom-up approach for either providing Internet access to unprivileged populations [5], [6] or for providing local applications and services as an alternative to the Internet [7], [8]. Even big companies like Microsoft in 2004 [9] investigated and proposed off-grid Wi-Fi mesh networks to provide ubiquitous connectivity to people at the neighborhood level.

In recent years, official stakeholders like the Internet Society (ISOC) and the Internet Engineering Task Force (IETF) introduce community networks and off-grid networking as a valid approach to connect the 4 billion yet unconnected all over the world [10]–[12].

Underserved rural areas remain one of the least attractive for investment by Internet Service Providers (ISPs) but at the same time they offer a great opportunity for off-grid networking solutions that will boost their potential in agriculture and local industry, but also solve their issues in healthcare, education and governance [13]. Others claim that "It's time to move from Broadband to Infrastructure" in order to assure network neutrality and unlock a more innovative business model [14].

However, current off-grid solutions struggle to guarantee transparency, decentralization and trust among the network participants using legacy protocols and most of the time centralized Internet protocols for services like identity management and data storage. In order to fill this gap, we designed a scalable architecture and leveraged open-source, state-of-the-art software for delivering an integrated solution that issues identities and stores data in a decentralized way.

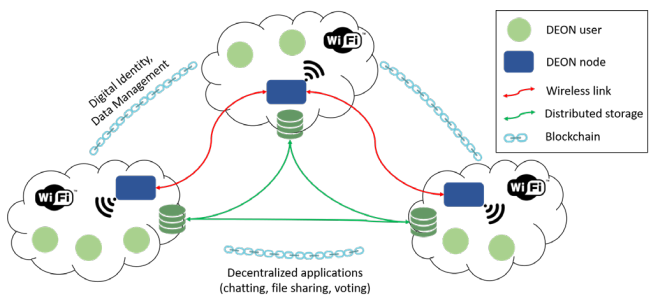

Figure 1: A Decentralized Off-grid Network

We address the challenges that pertain to identity and data management and apply the concept on an already established off-grid communication device called MAZI [15], [16]. MAZI enables impromptu network formation among users through peer-to-peer communication links. A MAZI node is a Raspberry Pi to which people can connect using their Wi-Fi devices and access services, even in settings with no access to the Internet. We augment this open-source toolkit with blockchain and decentralized technologies to enable the deployment of an open, decentralized and trusted data sharing infrastructure.

In summary, our main contribution is the design and implementation of a blockchain-based platform for deploying DEcentralized Off-grid Networks (DEON). More specifically:

- We design and develop a data management layer for transparent exchange of data across a distributed network in a secure and private manner.
- We design and develop an identity management layer for enabling network participants to control access to their identity data and be self-sovereign, namely be in control themselves over who gains access to their personal information and for what purposes.
- We combine Hyperledger Fabric, Hyperledger Indy and Hyperledger Aries towards an integrated solution for permissioned blockchain networks.
- We install the developed platform in a network of single-board computers (Raspberry Pi) and evaluate its performance in off-grid settings. The results show reasonable throughput and latency that can empower a plethora of off-grid use cases.

The rest of this paper is organized as follows. Section II gives a background of the MAZI open-source toolkit, the blockchain frameworks we use and the distributed file system. In Section III, we present the architecture of the DEON platform, and subsequently in Section IV we describe in detail the DEON components. Section V presents the performance evaluation of the platform, and Section VI describes related work. Section VII concludes with directions for future work.

## II. BACKGROUND

The inspiration for the acronym DEON comes from the Greek word "ΔEON," which is the root of the English word "deontology," and means something that is due and appropriate, a duty, an obligation. The metaphor is that our DEON platform gives - as is due - users off-the-grid access to their right to connect with each other and share data and services in a way that respects the principles of privacy, decentralization, user-centric data and self-sovereign identity.

### A. Network Infrastructure

DEON's network infrastructure is based on the MAZI open-source toolkit, a concrete set of hardware components, open-source software, artefacts and guidelines, which enables citizens to deploy their own off-grid networks, *MAZI Zones*. The hardware platform employed is the Raspberry Pi, optionally accompanied by off-the-shelf equipment like Wi-Fi USB adapters, Wi-Fi routers and USB disks, which enhance the provided functionalities of a MAZI zone. The MAZI software is a set of graphical interfaces and services intended for use by diverse groups of people, from network administrators to urban activists and artists. The user-friendliness of the interfaces enables the effortless configuration of the employed hardware and the management of various services, such as the configuration of the Wi-Fi Access Point parameters, the setup of a wireless mesh network, the collection of sensor measurements and the sharing of files.

The toolkit is following the DIY concept, allowing the user to select, buy and assemble low-cost equipment, load the MAZI software and consequently deploy one or more MAZI nodes in the desired area. A MAZI node exposes a Wi-Fi network that provides local applications and services. In addition, multiple MAZI nodes can be deployed in proximity connecting with each other and forming a Wi-Fi mesh network that expands the coverage area and brings closer people from distant neighborhoods.

The process of deploying and maintaining a MAZI Zone is comprehensively documented, targeting technology-savvy users as well as users without any technological background. This has stimulated adoption by communities for various scenarios; MAZI counts nearly 50 known deployments in Europe, Africa and South America serving hundreds of users [17]. Apart from the reported MAZI Zones, there are other unreported deployments operating either online or offline that will never be known to the public since this is the choice of the community operating them.

However, although the MAZI toolkit is offering a decent pool of applications and services operating in an off-grid network, the data and identity layers of a MAZI Zone are not distributed, but instead there is a super node that hosts all the databases and identities of the network. This can cause serious complications in such networks where transparency and security are essential ingredients to establish trust among the participants.

### B. Blockchains

An off-grid network is a distributed system, and thus inherits all of the distributed systems' characteristics and advantages, but also their challenges and limitations. One of the key challenges in such systems is the mechanism for reaching consensus among the nodes for concepts like file ownership, identities and data storage. A blockchain can address these challenges by providing a transparent method of communication, storing data and issuing identities.

Public permissionless blockchains are (aiming to be) perfectly decentralized and secure, but they require a lot of resources to operate and do not scale efficiently. Private permissioned blockchains sacrifice some of their decentralization, but they are scalable and at the same time can be lightweight and secure. We are using a private permissioned approach of combining multiple ledgers, each one for a specific purpose. A permissioned blockchain can establish the foundation for building trust in an off-grid network, where digital identi-ties and data exchange methods are managed by distributed ledgers, smart contracts and tamper-proof storage.

One of the most advanced initiatives for developing permissioned blockchains is the Hyperledger Project [18], which is hosted by the Linux Foundation and provides an umbrella for several enterprise-grade, open-source distributed ledger frameworks and associated codebases. Its goal is to provide the infrastructure for the development and operation of robust, industry-specific applications, platforms and hardware systems to support cross-organizational transactions. Currently, Hyperledger hosts several distinct projects going beyond distributed ledger frameworks, to include smart contract engines, client libraries, graphical interfaces, utility libraries and sample applications. Hyperledger technologies have been used commercially for pilots in full operational use in several sectors, including supply chain, healthcare, education, IoT, entertainment and financial services. We utilize three projects of the Hyperledger ecosystem and build on them to implement core functionalities of our system. These are Hyperledger Fabric, Hyperledger Indy and Hyperledger Aries.

Hyperledger Fabric [19] is one of the most popular and adopted blockchain frameworks. The codebase was initially open-sourced by IBM in July 2017 and is currently maintained by a large community of over 300 developers, 45 companies and 100+ individuals, who managed to release an LTS (Long Term Support) version 1.4 in January 2019. Fabric provides a modular architecture for execution of smart contracts (called chaincode), pluggable consensus and an identity management service. The Fabric network consists of *Peer Nodes*, which execute the smart contracts and host the ledger, an *Orderer Service*, which ensures the consistency of the blockchain, orders the blocks and distributes them back to the Peer Nodes, and an Identity management Service (called *Membership Service Provider* or MSP), which handles the identities of the network components and the users, using X.509 certificates issued by a *Certificate Authority* (CA).

Two of the biggest advantages of Fabric is that it uses standard, general-purpose programming languages (Python, Go, Java, Node.js) instead of blockchain-specific languages (e.g., Solidity for Ethereum) and does not rely on a cryptocurrency, although it can support one. In terms of performance, it can achieve end-to-end throughput of up to 1k transactions per second per channel and scales well to over 100 peers, depending on the network parameters [20]. In addition, there are recent studies that claim reaching throughput of up to 20000 transactions per second in a Fabric network after certain architectural changes [21].

We chose Fabric as our data logistics ledger to store pointers to our data, which are stored in the InterPlanetary File System (IPFS). Furthermore, we leverage Fabric's chaincode functionality, which together with a Decentralized Identity management system incorporates trust in the transactions happening in the off-grid network and enables the development of the business logic for diverse applications.

Indy [22] gathers experts from the Internet identity standards community, whose shared passion for a privacy-focused design of an Internet-wide identity layer establish it as the ultimate tool for self-sovereign identity. The main goal of Indy is to support independent identity rooted on a distributed ledger and contribute tools to other blockchain frameworks for providing interoperable digital identities. Sovrin Foundation [23] operates a public Indy instance, where a network of over 40 nodes (called *Stewards*) maintains a public-permissioned ledger allowing anyone to read and transact with it, but only Stewards to write to it. Only public data are stored on the ledger, like public keys and credential definitions, while private data are stored off-ledger in users' premises, enabling secure peer-to-peer interactions.

The Aries project [24] is the client-side software of Indy's code repository, namely SDK and agents, which was disaggregated and moved to a new Hyperledger project in order to provide blockchain-agnostic components to the community.

We use the Indy and Aries frameworks and by extension *Decentralized Identifiers* (DIDs), which are a new type of identifier for verifiable, "self-sovereign" digital identity. They are fully under the control of the DID subject and eliminate the need for a centralized authority or identity provider, which could be either absent or inaccessible in a completely offline setting of such networks. In addition, we give control of identity data back to the users of our platform by keeping private data at the devices of the users.

*C. IPFS*

IPFS [25] is a peer-to-peer distributed file system with increased mainstream adoption aspiring to be the file system of the Web 3.0 and re-decentralize the way the Internet operates. It is a content-addressable network that combines successful protocols from other peer-to-peer systems, but also evolves them providing a single cohesive system. It uses Distributed Hash Tables (DHTs) as a lookup service and as a routing table to find the stored data, a BitTorrent-inspired protocol to exchange the data and a content-addressable way for storing data inspired by Git's Merkle DAG. Although it makes possible to distribute high volumes of static data with high performance, it lacks support for dynamic data such as modern websites. Thus it makes a perfect match with a blockchain, serving the actual files in IPFS and storing the IPFS Content Identifiers (CIDs) on a ledger in order to timestamp and secure them as well as to keep them updated and enable data discovery.

III. SYSTEM ARCHITECTURE

In this section we introduce the DEON architecture and the DEON core processes for data and identity management, and subsequently explain the transaction flow.

*A. DEON Overview*

We use state-of-the-art Web 3.0 technologies in order to provide an integrated platform for deploying scalable networks for off-grid, transparent communications between individuals and things. Individuals, communities, or enterprises can utilise the platform to design, deploy and manage their own communication infrastructure that will operate in a fully decentralized manner, guaranteeing security and transparency and eliminating the need for any super node or other external Trusted Third Party (TTP). The main components of DEON are summarized in the following:

- *An off-grid communication toolkit* that provides the mechanisms to deploy off-grid networks offering local applications and services.
- *A general purpose ledger* for storing data logistics to allow data privacy, efficient data discovery and subsequent auditing.
- A ledger specifically designed for allowing *decentralized identity management* of user identities.
- A *distributed file system* for fast data discovery and resilient access to data, independent of low latency or connectivity to the Internet.

An off-grid communication device provides the network infrastructure of the system, supporting the deployment of a Wi-Fi mesh network, where users connect and have access to local, decentralized applications. Users utilize DIDs to authenticate themselves in any of their interactions with a DEON application or another network user.

Data and identity management are performed through the usage of distributed ledgers (Fabric and Indy) which DEON nodes host. The purpose of using the distributed ledgers is twofold: on the one hand, to store pointers to real data hosted by a distributed file system (IPFS), and on the other hand to support self-sovereign identities and verifiable credentials. A sketch of the architecture is shown in Fig. 2. The components are explained in detail in the sequel.

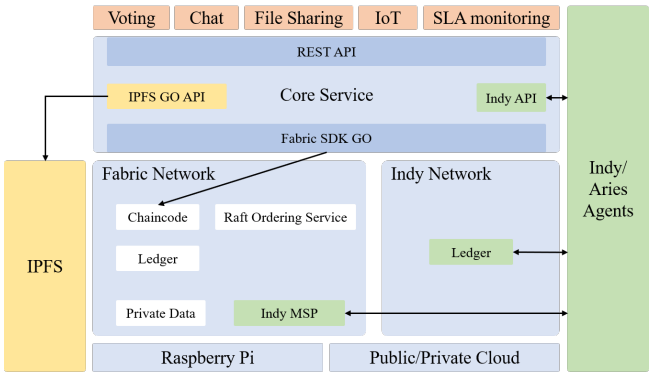

Figure 2: DEON software architecture

### B. Applications Marketplace

DEON enables a marketplace of applications which vary a lot, from social applications like file sharing, chatting, and digital voting, to IoT applications like anomaly detection in sensor measurements, and more enterprise-driven applications like contract signing and service level agreement monitoring. These applications use the data management layer to store data logistics and the identity layer for authentication and authorization of their users.

Applications in DEON are decentralized, taking advantage of the underlying decentralized infrastructure that is shaped from the deployment of distributed ledgers. Their front-end is stored in the distributed file system, while they use the DEON Core Service as shown in Fig. 3 to store transactions, sign transactions, authenticate an identity that wants to make an action in the network or discover data. Applications use also DIDs to authenticate themshelves and sign their transactions enabling the introduction of new applications in the network and their decentralized, on-the-fly registration and authentication.

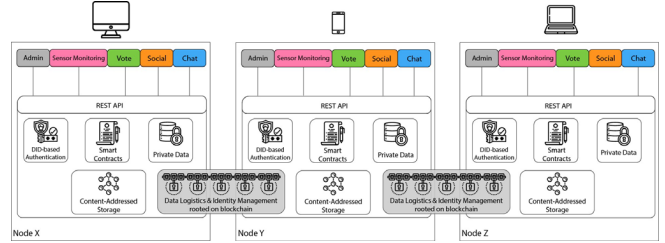

Figure 3: DEON Decentralized Marketplace

### C. Distribution and storage of data in DEON

DEON manages data coming from applications by leveraging both Fabric's ledger and IPFS. Applications send data to the Core Service which translates them to IPFS commands to store the actual data and to Fabric transactions to store data logistics and metadata. In this way, DEON eliminates central servers and promises secure peer-to-peer exchange of data with great performance. By removing super nodes, data are not owned or controlled by any of the nodes, bringing fairness and openness to the network. IPFS manages the data distribution across the network by storing file objects and DHTs and pushing files to users if and only if they ask for them. In addition, as a distributed file system, IPFS brings resiliency even without Internet access, keeping the network alive in case of a node disconnection or node failure.

On the data logistics part, Fabric's ledger stores pointers (hashes) to the actual data in order to enable later their discovery from other nodes and the auditing of the transaction history across the network. The pointers act as evidence that a transaction is made and are stored on Fabric's ledger, so that they can be accessed by any node. Furthermore we use Fabric's private data mechanism [26] to guarantee that only authorized nodes are able to retrieve sensitive data from IPFS (more details in Section III-E).

### D. Identities in DEON

We replace Fabric's native approach for user identities with a decentralized identity architecture that uses Zero-Knowledge Proofs, Verifiable Credentials (VCs) and pairwise DIDs, enabling self-sovereign identities. This is accomplished by leveraging the Indy and Aries frameworks and their privacy-by-design tools (see Section II-B). We follow a similar approach as in DEON's data management layer, where no private data are stored on-chain, but only pointers to them.

The actual identity data are stored in digital wallets, hosted by users' mobile devices (smartphones, tablets, laptops) which act like password managers and store Personal Identifiable Information (PII), such as VCs, DIDs and private keys. Upon their connection to the Wi-Fi network of one of the DEON mesh nodes, users get a DID and a VC which are stored in their private wallet for proving that they are members of the network. On the applications side and as described before, DEON applications send data transactions to the Core Service paired with their DID which are then signed by Aries agents and are forwarded to the Fabric network for verification and commitment.

The management of the aforementioned identity data within the DEON architecture is handled by Aries agents which connect to the local Indy ledger to issue DIDs and VCs to both users and applications, sign transactions, ask for users' connection endpoints and validate transactions' signers by verifying their proofs. By the VC-DID-based user authentication we verify that the user is a valid member of the network and he has the rights to use a specific application, and by the VC-DID-based transaction verification, we verify that the transaction is coming from a node of the network. The former eliminates adversarial users while the latter eliminates adversarial applications and nodes.

Finally, DEON DIDs and VCs - inheriting Indy's characteristics - are interoperable across local applications and services, the public Sovrin network and other Decentralized Identity systems as well, following the standards defined in W3C [27] and DIF [28]. In addition, network users are able to interact in a direct peer-to-peer manner through their agents and selectively disclose information during their interactions with other users as well as with the platform.

### E. Transaction flow

In the following, we describe the flow of a voting transaction invoked by one of the applications of the DEON marketplace, introduced in Section III-B. The specific transaction's chaincode[1] uses Fabric private data collections, but a similar approach applies for any other transaction that could happen within a DEON network, e.g., file transaction, sensor transaction, legal transaction or even a coin transaction.

*1) Pushing a vote:* As shown in Fig. 4, the invocation of a voting transaction consists of five phases. First, the vote is sent to the Core Service by a local application coupled with the application's DID and vote metadata. Then, the Core Service signs the transaction, pushes the data to IPFS and gets back the CID (*ipfs_hash*) representing the IPFS object. In the third phase, the *ipfs_hash* is pushed to the chaincode concatenated with a random value (salt) which will be used by the Fabric peer to create a hash which will be stored on the ledger (*public hash*). The fourth phase is the verification of the application's identity by the peer's MSP. During the fifth phase, the chaincode pushes the *public_hash* (*public_hash = hash {salt + ipfs_hash}*) to the Fabric ledger and the values of the salt and *ipfs_hash* to the private databases of the peers.

---
[1]DEON chaincode using private data collections: https://github.com/off-grid-block/off-grid-cc/tree/master/vote

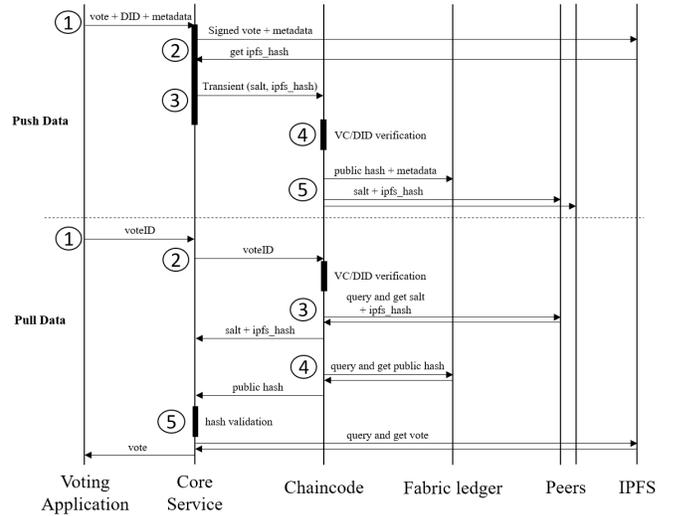

Figure 4: Transaction flow in DEON

*2) Querying a vote:* An application can query the DEON platform for a vote using the voteID (*voteID = pollID + voterID*). The Core Service queries the chaincode and, after the identity verification, the chaincode gets the salt and the *ipfs hash* of the vote from the local peer's private database. Then, the Core Service queries again the chaincode to get the public hash of the vote from Fabric's ledger. In this way, the public hash is compared against the hash from the peer's private database in order to verify that it has not been compromised. Finally, the Core Service gets the actual vote data from IPFS using the *ipfs hash* and forwards it to the application.

## IV. DEON COMPONENTS

In this section we describe in more detail each of the DEON components. shown in Fig. 5.

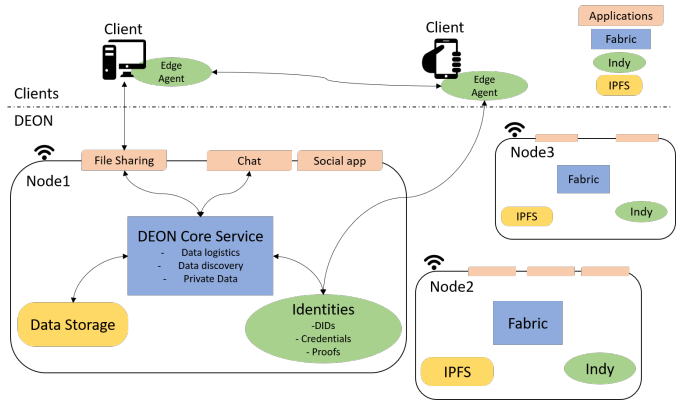

Figure 5: DEON software components

### A. Core Service and Data Storage

At the heart of the platform lies the Core Service, which acts as a middleware and joins the identity management components with the data logistics ledger and the distributed file system, while it also exposes all the functionalities to the application level. It is hosted by all the network nodes and it uses a Fabric client to interact with the local Fabric ledger and enable data discovery and data access control.

The Core Service is private-by-design and it ensures that no private data like votes, files, chat messages, or sensor measurements are stored on-chain, while it also guarantees the discovery of data by legitimate users. It exposes all its complex functionalities through a REST API, which enables applications to send data transactions coupled with user DIDs that are being forwarded to the underneath components for authentication and access control.

The Core Service takes advantage of the extensions in Fabric SDK Go described in Section IV-B that enable support for DIDs and the connection with an Aries Agent, namely an aca-py agent [29]. It calls the aca-py agent for actions such as getting a signing DID or signing transactions with DIDs.

The connection with IPFS is achieved through the IPFS Go API, which is used to push or retrieve data from the distributed file system. The CIDs acquired from IPFS are stored either in Fabric's private data collections in case of sensitive data or on Fabric's ledger in case of non-sensitive data. As a complementary security measure in order to prevent dictionary attacks, we hash the CID of sensitive data coming from IPFS appended with a random "salt" before storing them in private data collections. Metadata attributes from transactions are also stored on-chain, as they will later facilitate the discovery of the data by applications of other nodes and will enable the collection of usage statistics.

### B. Identity Management

Decentralized identity data are integrated in Fabric by extending the functionality of Fabric components, like the chaincode interface and the Fabric SDK Go, and by developing a custom MSP for the verification of DIDs. Fig. 6 shows the architecture of a DEON node - from the perspective of managing identities - which eliminates the need for Certificate Authorities.

More specifically, a custom Fabric MSP – the "Indy MSP" – is developed to verify and attest Indy's DIDs, Verifiable Credentials and Proofs. Similar approaches have already been followed by the Fabric community for implementing other custom MSPs (e.g., an Idemix MSP [30]). Regarding the chaincode, we extended the client identity chaincode library to be able to retrieve and manage attributes from an Indy DID in order to make access control decisions based on them. Finally, the Fabric SDK has been extended to support provision of DIDs and signing of transactions based on them.

The distributed ledger for enabling the decentralized identities is the Indy ledger, which is a public (within the DEON network) permissioned one, is deployed in all DEON nodes and hosts identity records.

For the communication between users or other interactions involving DIDs, we leverage existing Aries agents to serve as clients of decentralized identities. We deploy an enterprise/cloud Aries agent, application agents and mobile edge agents. The cloud agent is installed on DEON nodes and plays the role of the issuer of DIDs and VCs as well the role of the verifier/MSP of DID-based Fabric transactions. The application agents are the identity managers of applications and are used to manage application DIDs and sign transactions. The edge agents are installed on users' devices and enable users to store and manage and share their identity data.

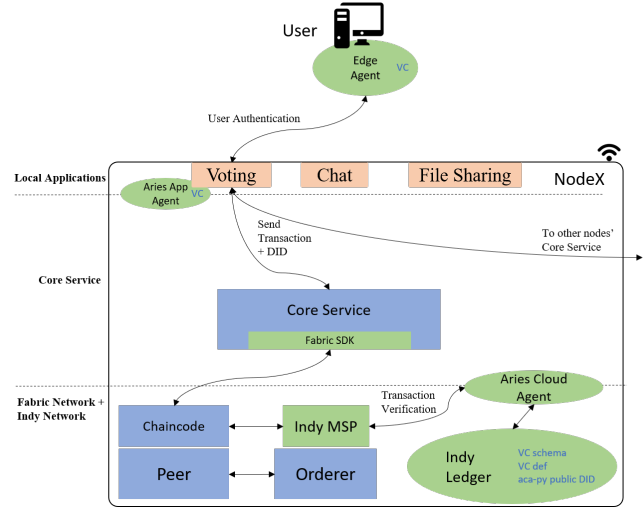

Figure 6: Technical architecture of the integration of Fabric and Indy

## V. EVALUATION

This section presents a performance evaluation of the proposed platform in "off-grid settings," namely an offline, local network of Raspberry pi boards, representing DEON nodes, where the DEON software components were installed.

### A. Experiment Setup

We used three Raspberry pi 3 B+ boards[2] interconnected through a 1 Gbit/s switch and installed an unofficial Ubuntu 18.04.3 version[3] which provides the 64-bit environment we need to run the Fabric code. The SD card we used in all Rpis is the Lexar High-Performance 633x 32GB microSDHC U1. In addition, we had to build the Docker images for Fabric v1.4.2 (fabric-tools, fabric-ccenv, fabric-orderer, fabric-peer, fabric-couchdb) using the fabric-base images for arm64, since we did not find any of these available online. The Docker images produced are publicly available in Docker hub[4].

Each Rpi represented a different Fabric organization consisting of one endorsing peer and using couchdb for its peer state database. We used the RAFT-based ordering service and each Rpi node hosted an orderer to ensure decentralization and robustness in case of a node failure. All peers were connected under a single Fabric channel and the policy for our chaincode required a signature from each of the organizations, since we need to ensure that transactions are validated from all the DEON nodes.

---
[2]Raspberry Pi 3 Model B+: https://www.raspberrypi.org/products/raspberry-pi-3-model-b-plus
[3]Ubuntu Server 18.04.3 for Rpi https://github.com/TheRemote/Ubuntu-Server-raspi4-unofficial/releases
[4]Docker images for Fabric v1.4.2 in arm64 architecture https://hub.docker.com/u/haniavis

The client was hosted by a server equipped with Intel Core i7-8650U at 1.9 GHz, 16 GB of RAM and an SSD for storage, and it generated 1000 parallel transactions in varied rates towards all the nodes' peers, using the latest release of the Fabric SDK Go[5].

### B. Analysis of the results

*1) Goal:* Our goal is to examine the performance of the DEON platform running on devices of limited computational power and most specifically on Rpis. We assessed the platform as a whole, as well as without the part of pushing data to IPFS, since we want to share with the research community performance measurements of the latest stable Fabric version (v1.4) in arm64 architecture. The impact of using DIDs on the performance of the system is left as future work, since we use Fabric's native certificates in our experiments. Moreover, we tested two different chaincodes, with and without using Fabric's private data concept, to evaluate this recent addition to the Fabric code.

*2) Results:* We made end-to-end experiments measuring throughput (transactions per second - tps) and latency for pushing data to the DEON platform. The client sent transaction traffic in rates ranging from 50 tx/sec to 200 tx/sec and we measured time for transaction approval according to already defined procedures by the Hyperledger community [31]. Fig. 7a, 7c show the throughput and latency achieved for different block sizes and without Fabric's private data, including or not the calls to IPFS. Subsequently, Fig. 7b, 7d show throughput and latency achieved with the private data chaincode.

Overall, we saw acceptable performance for diverse applications, reaching up to 100 tps throughput and sub-second latency in distinct experiments. However, the results show significant overhead on the performance of the platform introduced by the calls to IPFS and the use of private data in Fabric chaincode. This is the cost of securing the data management layer and can be tolerated in specific applications like voting, where latency is not crucial while security and transparency are imperative.

## VI. Related Work

### A. Off-grid communication

Several proposals related to off-grid networking exist, either commercial or community-driven. GoTenna [32], Beartooth [33], and Sonnet [34], [35] offer mesh networking technology intended to replace walkie-talkie devices and to extend their functionality, by connecting a small physical device to the user's smartphone and enabling multimedia messaging, location sharing, and offline maps. The mesh network created by Sonnet devices can additionally be connected to the Internet

---

[5]Fabric SDK Go v1.0.0-beta1 https://github.com/hyperledger/fabric-sdk-go/releases/tag/v1.0.0-beta1

---

as long as one node is online. All are proprietary, focus on long range versus throughput, and use encryption to secure communication, but exhibit limited functionality compared to our platform. While encryption suffices for securing messaging applications, it is not enough for more elaborate ones that involve access control. DEON, in addition to enabling meshing (albeit in shorter ranges), surpasses this obstacle via using the advanced concept of DIDs.

LibreMesh [36] is "a modular framework for creating OpenWrt-based firmware for wireless mesh nodes." Its firmware allows "simple deployment of auto-configurable, yet versatile, multi-radio mesh networks," and it uses a dynamic routing protocol which makes the entire mesh network look like a single LAN to the user. However, it only provides networking infrastructure and no distributed applications, storage or access control.

Rightmesh [37] is a software-based mobile meshnetworking approach that uses smartphone devices connected in a serverless infrastructure. A token is used for incentivization; online users can share their Internet connection with off-grid users by exchanging tokens. A key difference with DEON is that Rightmesh uses Ethereum as the underlying blockchain, and only for the token's purposes, as opposed to DEON, which stores also data logistics and DIDs on distributed ledgers. Moreover, Rightmesh implies the installation of an Android application on the user's device, while DEON's operation is browser-based and thus non-invasive. Finally, for the token to be used in the Rightmesh network, at least in the beginning some "superpeers" need to be deployed, which makes the architecture depart from the totally decentralized model by introducing the need for trusting whoever is in control of those nodes. On the contrary, DEON is totally decentralized, with no nodes having extra privileges.

An analog of our platform that uses the current Internet infrastructure instead of operating off-grid is the SAFE (Secure Access For Everyone) Network [38]. It joins together the spare computing resources of its users, creating a global network and giving control of the data back to them. The software is available to download for free and allows for custom distributed application development. Users are rewarded with a token called "Safecoin" for sharing their hardware and network resources. SAFE extends the Kademlia distributed hashtable, which is also the one used internally by IPFS, and data are encrypted on each user's computer before they are fragmented and spread throughout the network. The effort is still in development. We note that while DEON can support a cryptocurrency if needed for a specific application, it is not required for the proper function of the platform (i.e. for the consensus).

### B. Platforms for Decentralized Data Sharing and Computation

The work closest to ours is [39], where the authors describe a framework for data ownership, data transparency and auditability, and fine-grained access control. Their network consists of users, services, and nodes (nodes maintain the

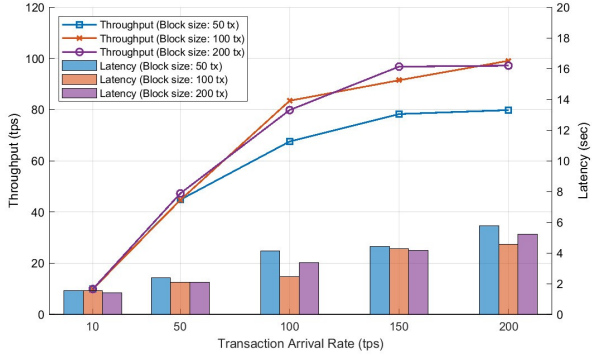 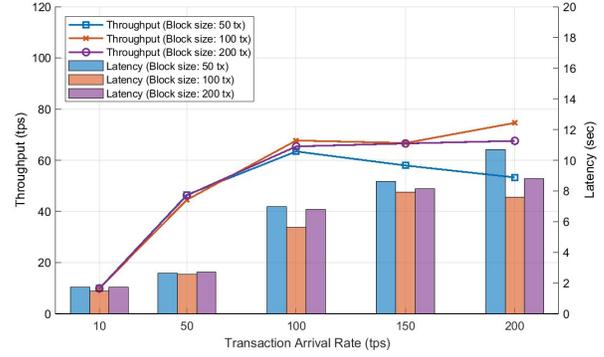
(a) No IPFS, no private data  (b) No IPFS, with private data

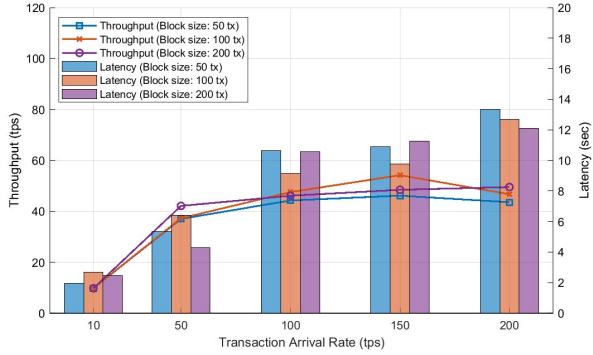 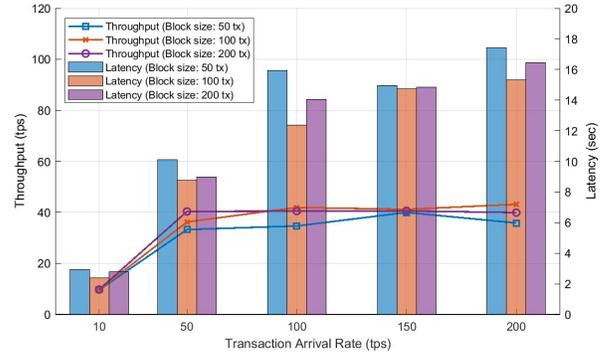
(c) With IPFS, no private data  (d) With IPFS, with private data

Figure 7: Throughput and latency of DEON as a function of the transaction arrival rate for different system configurations

blockchain). They use signatures for identification of users, and a single blockchain for both access control and data transactions, with the real data distributed and replicated in an off-chain Kademlia-based key-value store, with a role similar to the one IPFS plays in our solution. The authors define their own identity and access control mechanisms, and in addition propose Multi-Party Computation (MPC) over encrypted data, which we do not consider.

The Oasis platform [40] wants to provide the infrastructure for a decentralized cloud, offering trustless privacy. DFINITY [41] is a "virtual blockchain computer" running on top of a peer-to-peer network. Despite the similarities, both these platforms operate over the Internet and not off-grid, so their focus is different than DEON's.

Selimi et al. [42], motivated by the issue of participation incentivization in community mesh networks like Guifi [43] – with proper compensation for services offered – observe that this was being done in a centralized manner, and attempt to decentralize this procedure by deploying Hyperledger Fabric over Guifi. They do not take identities into consideration though, and their description is more at a proof-of-concept level, as they only use a single Fabric organization.

## VII. Conclusion & Future Work

In this paper we introduced DEON, a decentralized data sharing infrastructure for deploying privacy-respecting, decentralized applications in off-grid settings, where no access to the Internet is available. We designed and implemented a data management layer for maintaining the transparency of data exchanges and an identity management layer for giving the users privacy and self-sovereignty of their identity data, and combined these with a decentralized file system and open-source off-grid devices that create a mesh network. As a future step, we intend to extend the use of DIDs to Fabric network components (peers, orderers, MSPs) to further eliminate single points of failure. We also intend to study whether the local Indy ledger can be removed from the nodes and use the public ledger (run by Sovrin) instead. Furthermore, we hope to further improve DEON's performance by experimenting with other types of Fabric's state database (e.g., leveldb) or different chaincode policies. In any case, our platform is a step towards the realization of the Web 3.0 and the decentralization of our vulnerable centralized infrastructure via secure privacy-focused distributed alternatives.

## Acknowledgements

This work is supported by NSF 1932220 and Tata Consultancy Services (TCS). In particular, we thank our collaborators in TCS for their contribution and insights, especially on the identity management component of our platform.